\documentclass[12pt,a4paper]{article}
\usepackage{amsmath}
\usepackage{graphicx}
\usepackage{psfrag}

\title{Finite size scaling of the balls in boxes model}

\author{P.~Bialas${}^{a,b}$, L.~Bogacz${}^{c}$,
Z.~Burda${}^{c,d}$ and D.~Johnston${}^e$ \\ \\
\small ${}^a$ Institute of Comp.~Science, 
Jagellonian University, 30-072 Krak\'ow, Poland\\
\small ${}^b$Fakult\"at f\"ur Physik, 
Universit\"at Bielefeld, 33501 Bielefeld Germany \\
\small ${}^c$Institute of Physics, Jagellonian University, 
30-059 Krak\'{o}w, Poland\\
\small ${}^d$ Laboratoire de Physique Theorique\thanks{Unit\'{e} Mixte 
de Recherche 8627} \\
\small Universit\'{e} de Paris Sud, 91405 Orsay Cedex, France \\
\small ${}^e$ Dept. of Mathematics, Heriot--Watt University,\\\small 
Edinburgh EH14 4AS, Scotland }

\date{}
\begin{document}
\maketitle

\begin{abstract}
We discuss the finite size behaviour in the canonical ensemble of the
balls in boxes model. We compare theoretical predictions and numerical
results for the finite size scaling of cumulants of the energy
distribution in the canonical ensemble and perform a detailed
analysis of the first and third order phase transitions which appear
for different parameter values in the model.
\end{abstract}

\section*{Introduction}

The balls in boxes model is a simple statistical model describing
an ensemble of balls distributed in boxes subject only to a single global
constraint.
In some circumstances the model undergoes a phase transition 
driven by a condensation of the balls into  a single box. 

The model has been introduced as a mean field approximation
to lattice gravity \cite{bbpt,bbj1}. Despite its simplicity,
the model captures the main features of the phase transition observed
in lattice Euclidean quantum gravity models 
({\em ie} dynamical triangulations) such as
the discontinuity of the transition \cite{bbkp,db} 
and the appearance of singular structures \cite{hin1,hin2,rck,ckrt}. 

Variations of the phase transition of the balls in boxes type
can be found in  many areas of physics
like the thermodynamics of hadrons or strings  \cite{hag,ven},
branched polymers \cite{bb2,jk} or percolation\footnote{The appearance
of a percolating cluster in the percolation theory
can be translated into the condensation 
of balls in one box, which after condensation contains extensive 
number of balls.}. The balls in boxes model is 
also closely related to urn models \cite{u1,u2}
and the spherical model \cite{bk}.
It is thus a rather generic type of phase transition.

The model has an interesting phase structure~: by tuning a single
parameter one can change the order of the transition or make the
transition disappear entirely \cite{bbj1,bbj2}.  This sort of phase
transition was discovered in the spherical model by Berlin and Kac
\cite{bk}.  The nature of the transition is quite different from 
more standard
phase transitions in the theory of critical phenomena
which result from the appearance of long range spatial correlations.
Here the transition does not refer to any
correlations in space. Instead, it is a kinematic condensation which
comes about when a fugacity of the series representing the partition
function hits the radius of convergence of the series.  This fugacity
sticks to the radius of convergence and cannot move.  At this point
the corresponding physical system enters the condensed phase. From
this point of view the mechanism of the transition is very similar to
Bose--Einstein condensation. The basic difference between the balls in
boxes type of condensation and the Bose-Einstein one is that in the
latter the system condenses into an energetically favoured ground
state, while in the former balls condense into a box which is
indistinguishable from the remaining ones. The balls in box
condensation thus spontaneously breaks the permutation symmetry of the
boxes.

The thermodynamic limit of the balls in boxes model and the relation
to lattice gravity has been discussed in a series of papers
\cite{bbpt,bbj1,bbj2}.  Here we discuss finite size effects and confront
theoretical predictions with the numerical analysis. The numerical
analysis is done using an exact algorithm which recursively generates
partition functions for systems of moderate size {\em ie} up to a few
thousand boxes.

\section*{The model and thermodynamic limit}

In this section we briefly recall the model \cite{bbj1,bbj2}. 
The model describes $N$ balls distributed in $M$ boxes.
Each box has a certain weight $p(q)$ 
which depends only on the number $q$, of balls in it. 
The weights $p(q)$ are the same for all boxes. 
For a given partition of balls in boxes $\{ q_1,q_2,\dots,q_M\}$,
the total weight of the configuration is equal to the product
$p(q_1)\dots p(q_M)$ -- {\em ie} the boxes are almost independent. 
They are not entirely independent
because of the constraint $q_1 + \dots + q_M = N$. It is
this constraint which makes the model nontrivial.
The model is defined as a sum over all
partitions of $N$ balls in $M$ boxes weighted by
the product weight. This is a fixed density ensemble,
with density $\rho=N/M$. In the thermodynamic limit 
$M\rightarrow \infty$, and $\rho=const$, 
the behaviour of the system depends on the density $\rho$.
For some choices of weights $p(q)$, there exists a critical 
value $\rho_{cr}$ above which the condensation of balls takes 
place, which means that a box appears in the system which 
contains a finite fraction of the balls.
The phase transition in $\rho$ corresponding to this
condensation is of third or higher order \cite{bbj1}.

The fixed $\rho$ ensemble is a kind of microcanonical ensemble. One
can also consider ensembles with varying density. 
There are two
natural candidates for such ``canonical'' ensembles~: for a fixed $M$
one can allow for varying $N$ by introducing a conjugate coupling to
$N$.  In this formulation boxes decouple and we are left with $M$
copies of the urn model.  Alternatively for fixed $N$ one can allow
for varying $M$. This corresponds to the ensembles studied in lattice
gravity \cite{bbpt} and will be the subject of our investigations.

The partition function for this canonical ensemble reads~: 
\begin{eqnarray}
Z(N,\kappa) &=& \sum_{M=1}^{\infty} e^{\kappa M}
\sum_{q_1,\ldots,q_M} p(q_1)\cdots p(q_M) \ \delta_{q_1+\cdots+q_M,N} 
\\
&=& \sum_{M=1}^{\infty} e^{ \kappa M} z(N,M) \, .
\label{z}
\end{eqnarray}
The canonical function can formally be treated as a discrete Laplace
transform of the microcanonical partition function $z(N,M)$ as the
second line of the equation above shows.  Analogously, in the same
language, we can introduce the grand--canonical partition function as
the Laplace transform of the canonical one~:
\begin{equation}
{\cal Z}(\mu,\kappa) = \sum_{N=1}^{\infty} 
e^{-\mu N} Z(N,\kappa) \, .
\label{trans}
\end{equation}
Each box is assumed to contain at least one ball. Therefore
the sum in the last equation starts from $N=1$. 

The weights $p(q)$ are {\rm a priori} independent parameters
that we assume to be non--negative. 
In order that the large $N$ limit 
be well defined the weights cannot grow faster than exponentially 
with $q$. On the other hand the exponential growth factor is irrelevant
from the point of view of the critical behaviour of the model,
as can be seen below. Namely, if one changes weights as~:
\begin{equation}
p(q) \rightarrow p'(q) = e^{-\kappa_0} e^{\mu_0 q} p(q) \, ,
\label{k0mu0}
\end{equation}
where $\mu_0$ and $\kappa_0$ are some constants,
the partition functions change as~: 
\begin{eqnarray}
z(N,M) \rightarrow z'(N,M)
&=& e^{\mu_0 N-\kappa_0 M} z(N,M) \nonumber \\
Z(N,\kappa) \rightarrow 
Z'(N,\kappa) &=& e^{\mu_0 N} Z(N,\kappa-\kappa_0) \\
{\cal Z}(\mu,\kappa) \rightarrow
{\cal Z}'(\mu,\kappa) &=& 
{\cal Z}(\mu-\mu_0,\kappa-\kappa_0) \, .
\nonumber
\end{eqnarray}
Thus the change of the exponential factor in the weights merely causes 
a redefinition of the coupling constants 
$\mu$ and $\kappa$. 
So from the point of view of the phase structure of the
model, only the sub-exponential factors of weights matter.
Here we restrict ourselves to the power like weights~:
\begin{equation}
p(q) = q^{-\beta} \ , \quad q=1,2,\dots \, ,
\label{w}
\end{equation}
which capture a variety of essential phase transitions
in the model. Some remarks on the behaviour for weights 
of general form can be found in \cite{bbj2}.

In the large $N$ limit the model can be solved analytically \cite{bbj2}.
The free energy~:
\begin{equation}
\phi(\kappa)  = \lim_{N\rightarrow \infty}
\frac{1}{N} \log Z(N,\kappa)
\label{phi}
\end{equation}
has a singularity in $\kappa$ at $\kappa_{cr} = -\log \zeta(\beta)$,
for $\beta \in (1,\infty)$ and has no singularity otherwise.
$\zeta(\beta)$ is the Riemann Zeta function. 
The phase for $\kappa>\kappa_{cr}$ is called fluid,
and for $\kappa<\kappa_{cr}$ condensed.

The order parameter for the transition is the first derivative
of the free energy with respect to $\kappa$ which corresponds
to the inverse average density of balls per box~:
\begin{equation}
r = \phi^{(1)}(\kappa) 
= \frac{\partial \phi}{\partial \kappa} = \frac{\langle M \rangle}{N} \, .
\end{equation}
It vanishes in the condensed phase {\em ie} for 
$\kappa < \kappa_{cr}$. The average $\langle \dots \rangle$
is taken with respect to the canonical ensemble (\ref{z}).

The order of the transition depends on $\beta$. For $\beta$ in the
range $\beta \in (2,\infty)$ the transition is first order.  When
$\kappa \rightarrow \kappa_{cr}^+$ approaches the critical point the
order parameter goes to a nonzero constant
$r_{disc}=\zeta(\beta)/\zeta(\beta-1)$ corresponding to the
discontinuity of the order parameter at the critical point.  If one
treats $M$ in the partition function (\ref{z}) as the energy of the
system, then the discontinuity of $r$ corresponds to the latent
heat. For $\beta \rightarrow 2$ the latent heat vanishes and the
transition becomes continuous.  For $\beta$ in the range $\beta \in
(1,2]$, $r$ has a branch point singularity at $\kappa_{cr}$. In this
case, the order parameter vanishes when one approaches the critical
point $\Delta \kappa = \kappa - \kappa_{cr} \rightarrow 0^+$~:
\begin{equation}
r = \phi^{(1)} \sim \Delta \kappa^{x_1}, \makebox[2cm]{where}
x_1 = \frac{2-\beta}{\beta-1} \quad .
\label{x}
\end{equation}
The transition is $n$-th order for $\beta \in
(\frac{n+1}{n}, \frac{n}{n-1}]$. When $\beta$ approaches one ($\beta
\rightarrow 1^+$) the transition becomes softer and softer
($n\rightarrow \infty$) and eventually disappears at $\beta=1$. For
$\beta$ smaller than one the system has only
a fluid phase and displays no phase transition.

\section*{Finite size scaling}

Let us denote the most singular term in the free energy by
$\phi_{sing}(\kappa)$ and the related exponent by $x_0$~:
\begin{equation}\label{phi_sing}
\phi_{sing} \sim \Delta \kappa^{x_0} \, .
\end{equation}
To match it with the scaling (\ref{x}) for
$\beta \in (1,2)$, we must have~:
\begin{equation}
x_0 = \frac{1}{\beta-1} \, .
\label{x0}
\end{equation}

For large but finite $N$ one expects that this behaviour 
gets substituted by the double scaling law with two exponents
which we denote by $A_0$ and $B$(see \cite{b1983} and reference therein)~:
\begin{equation}
\phi_{sing}(\Delta\kappa, N) = N^{A_0} f(\Delta \kappa N^B) \, ,
\label{fss}
\end{equation}
where $f(\xi)$ is a certain universal function of one argument
$\xi = \Delta \kappa N^B$. 
When $\Delta \kappa$ is fixed and $N$ goes to infinity 
one expects to asymptotically recover the singularity of the 
thermodynamic limit 
$\phi_{sing}(\Delta\kappa, N\rightarrow \infty) \sim \Delta\kappa^{x_0}$
independent of $N$.
This means that the universal function must have the following 
large $\xi$ behaviour~: $f(\xi) \sim \xi^{x_0}$ and~:
\begin{equation}
x_0 = -A_0/B \, .
\end{equation}

The singular part of the free energy is mixed with less singular and
analytic parts. While for $\Delta\kappa\rightarrow0$ the
$\phi_{sing}$ will eventually dominate over less singular parts it
can still itself be dominated by the analytic part. It is therefore
difficult to extract the scaling part $\phi_{sing}$ directly from the
function $\phi$. Instead, one does it by calculating the $n$-th
derivative of $\phi$ for $n$ large enough that the exponent $x_n$ is
negative.  We use the convention that the exponent $x_n$ is the power
$\phi^{(n)}_{sing}(\kappa) \sim \Delta \kappa^{x_n}$ of the most
singular part of the $n$-th derivative. Thus we trivially have $x_n =
x_0 - n$. The first value of $n=1,2\dots$ for which $x_n$ is negative,
gives the order of the transition. For this $n$, the singular part of
the $n$-th derivative blows up at $\kappa_{cr}$  and dominates the
analytic part.  Therefore in this case one can skip the subscript
$sing$ when writing a scaling formula analogous to (\ref{fss}) for
divergent derivatives~:
\begin{equation}
\phi^{(n)}(\Delta\kappa, N) = N^{A_n} f^{(n)}(\Delta \kappa N^B) 
+ \dots \, ,\quad\text{with}\quad A_n=n B + A_0
\label{fssn}
\end{equation}
keeping in mind that there are corrections, (denoted by dots)
which may be important for finite $N$. Asymptotically, 
for $N\rightarrow \infty$ they are, however,
negligible in comparison to the displayed part.

The knowledge of the exponent $x_0$ does not suffice
to calculate $A_0$ and $B$. In the standard finite size
scaling analysis the value of the exponent $B$ can
be obtained by a simple argument \cite{ff1967,b1983}. 
There are two relevant length scales in a $d$--dimensional 
system undergoing a continuous phase transition~:
the linear extension $L=N^{1/d}$ and the correlation 
length $\xi$, diverging as
$\xi\sim (\Delta T)^{-\nu}$, when $\Delta T = T-T_{cr}$
goes to zero. The critical behaviour sets in when the
lengths become comparable~:
\begin{equation}\label{fisher_crit}	
L\sim \xi
\end{equation}
giving~:
\begin{equation}	
\Delta T \sim N^{-\frac{1}{d \nu}} \quad \longrightarrow \quad
B=\frac{1}{d\nu} \, .
\label{B}
\end{equation} 
Alternatively one can say that critical properties
of the system depend on the dimensionless ratio~: $\xi/L$.
We use here a similar reasoning. Namely, we extract
characteristic scales from the distribution of the
box occupancy number $\pi(q)$ {\em ie}
the probability that a box has $q$ balls~:
\begin{equation}	
\pi(q)=\bigg\langle \frac{1}{M}\sum_{i=1}^M \delta(q_i - q) \bigg\rangle
= \frac{p(q)e^{\kappa}Z(N-q,\kappa)}{Z(N,\kappa)} \, .
\end{equation}
In the large $N$ limit the distribution has the form~:
\begin{equation}	
\pi(q)\sim p(q)\,e^{-\phi(\Delta\kappa) q}
\end{equation} 
as can be seen from (\ref{phi}). 
There are two scales which govern this distribution
when the critical point is approached from the fluid phase~:
the damping scale factor $[ \phi(\Delta\kappa)]^{-1}$ and 
the system size $N$. 
Thus, in analogy to (\ref{fisher_crit}) one expects
that the dimensionless combination of the scales 
which defines universal critical properties of the system
is~: $ [ \phi(\Delta\kappa)]^{-1}/N$. 
Using the relation (\ref{phi_sing}) we eventually get~:
\begin{equation}
B=x_0^{-1}=\beta-1\quad\text{and}\quad A_0 = -1 \, .
\end{equation}
Inserting this into (\ref{fss}) we obtain $A_n = nB - 1$.  This is
a sort of Fisher scaling relation\footnote{In the standard
considerations of the second order phase transition in the theory of
critical phenomena the exponent $B= 1/\nu d$ and $A_2 = \alpha/\nu d$,
The exponent $\alpha$ is the heat capacity exponent {\em ie}
the second cumulant of energy. Thus, the relation $A_2 = 2B -1$
corresponds to $\alpha = 2 - \nu d$, known as Josephson's or Fisher's
law.}. 

There are some thermodynamic inequalities which follow directly from
the Fisher relation. Firstly, one can see that $B$ must be greater
than zero, $B>0$, in order that there exist $n$ such that
$A_n>0$. Otherwise there were no transition. This inequality is in
accordance with the inequality $\beta>1$ which gives the condition for
existence of the transition in this model, as discussed in the text
after the equation (\ref{x}). Secondly, $A_2 = 2B-1$ must be smaller
than one $A_2<1$ which amounts to $B\le 1$. This is because the second
cumulant may not grow faster than $\frac{1}{4}N$ as can be seen from
(\ref{f2}) by taking into account the fact that values of $M$ lie in
the range $[1,N]$.  In fact the limiting value $B=1$ corresponds to
the first order transition scaling which follows from the presence of
a non--vanishing latent heat\footnote{In the standard theory of
critical phenomena $B=1/\nu d = 1$ gives the canonical exponent
$\nu=1/d$.} {\em ie} $r_{disc}>0$.  This holds for all $\beta\le
2$. In this range of $\beta$ the exponents $A_2=B=1$ are constant,
while the discontinuity $r_{disc}$ changes with $\beta$.

Formula (\ref{fssn}) gives a practical way of computing the 
most relevant singularity of the partition 
function from the finite size analysis of numerical data. 
Namely, one computes divergent derivatives,
for them estimates the exponents $A_n,B$ and
using the relation $A_n = A_0 + nB$ one calculates~:
\begin{equation}
x_n = -A_n/B = x_0 + n \, .
\end{equation}
Numerically, derivatives $\phi^{(n)}$ of the free energy
are computed as cumulants of the canonical distribution for $M$ (\ref{z})~:
\begin{eqnarray}
\phi^{(1)} &=& \frac{\langle M \rangle}{N} \\
\phi^{(2)} &=& \frac{\langle M^2\rangle - \langle M\rangle^2}{N} 
\label{f2} \\
\phi^{(3)} &=& \frac{\langle M^3\rangle - 3\langle 
M^2\rangle \langle M \rangle +
2 \langle M\rangle^2}{N} \, .\\
\dots 
\nonumber
\end{eqnarray}
For finite $N$ the first cumulant $\phi^{(1)}(\kappa,N)$
grows monotonically from zero to one as a function of $\kappa$,
while the higher order cumulants $\phi^{(n)}$ have
$n-1$ extrema. If a cumulant is divergent,
the leftmost maximum corresponds to the scaling, critical
part described by the formula (\ref{fssn}). 
The remaining extrema lie in the fluid phase and come 
from the non-scaling part. In the thermodynamic limit the 
leftmost maximum of a divergent cumulant
approaches the singularity~: 
$\phi^{(n)}(\kappa) \sim \Delta \kappa^{x_n}$.

\begin{figure}
\begin{center}
\psfrag{512}[l][l][.75][0]{$512$}
\psfrag{1024}[l][l][.75][0]{$1024$}
\psfrag{Infty}[l][l][.75][0]{$\infty$}
\psfrag{kappa}{$\kappa$}
\includegraphics[clip,width=8cm,bb=77 65 510 430]{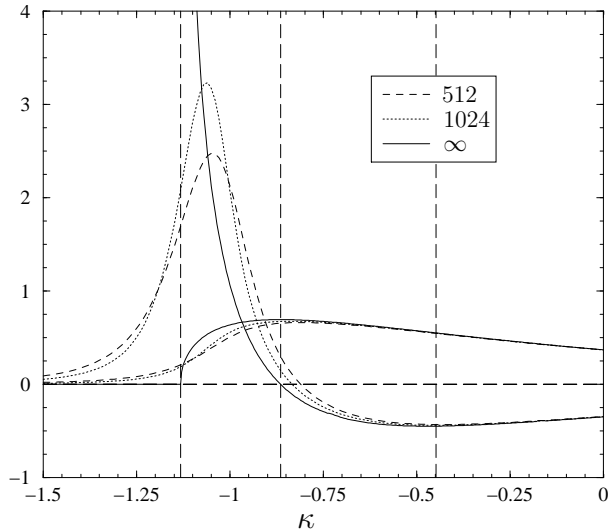}
\end{center}
\caption{\label{3deriv} 
The second and third derivatives of $\phi(\kappa,N)$ 
at $\beta=1.4$ in thermodynamical limit and for two different
system sizes. 
The critical value is at $\kappa_{cr}\approx -1.133$,
the position of the maximum of the second cumulant is 
at $\kappa_{max}\approx -0.865$, and  the minimum of the third one 
at $\kappa_{min}\approx -0.449$. Those are indicated by dashed
vertical lines.}
\end{figure}

As an example consider the 
figure~\ref{3deriv} where derivatives of $\phi(\kappa,N)$ are plotted
for $\beta=1.4$.  
In this case 
system undergoes third order phase transition at
$\kappa_{cr}=-\log \zeta(1.4)$. 
When $\kappa$ approaches the critical
value $\Delta \kappa = \kappa - \kappa_{cr} \rightarrow 0^+$
the first and second derivatives of $\phi(\kappa)$ vanish
as $\phi^{(1)} \sim \Delta \kappa^{3/2}$,  
$\phi^{(2)} \sim \Delta \kappa^{1/2}$,
whereas the third one diverges as
$\phi^{(3)} \sim \Delta \kappa^{-1/2}$. 
The second derivative has a maximum that lies in the
fluid phase far from the transition (\ref{3deriv}).
The third derivative has a minimum in the fluid phase
away from the critical value $\kappa_{cr}$.
For finite size the third derivative $\phi^{(3)}(\kappa,N)$
has a maximum and a minimum. When $N$ goes to infinity,
the position of the maximum tends to $\kappa_{cr}$ and its height grows 
according to a finite size scaling (\ref{fssn}). 
The position of the minimum approaches a value $\kappa_{min}$ which
lies in the fluid phase far from the critical region. 

For higher order cumulants the number of such noncritical extrema
increases with the order of the derivative. It may even happen that
for a given system size the height of a non critical maximum is larger
than of the critical one.  Generally, to determine $A_n$ and $B$ one
should analyze only the scaling of the leftmost maximum of a cumulant,
where the information about the singularity of $\phi^{(n)}$ is
encoded.

In the next sections we will present a numerical analysis
of the finite size data for the different ranges of $\beta$.
Before doing this, let us briefly describe the algorithm
to generate the finite size partition function is $Z(N,\kappa)$

We calculate the partition function $Z(N,\kappa)$ in two steps. 
First we compute values of the microcanonical partition function $z(N,M)$ 
for all $M$ in the range $[1,N]$ by the following recurrence 
relation:
\begin{equation}
z(N,M) = \sum_{q=1}^{N} z(N-q,M-1)p(q),
\label{recur}
\end{equation}
with the initial condition $z(1,q) = q^{-\beta}$.
Inserting the numbers $z(N,M)$ directly to the definition (\ref{z})
we obtain $Z(N,\kappa)$. The maximal size $N$ which can be reached
by this procedure is a few thousand. It is limited by floating point 
instabilities accumulated in the recurrence relation (\ref{recur}). 
To test the stability we check whether the results stay intact (up to
a shift in $\kappa$) under the change of weights (\ref{k0mu0}).
\begin{figure}
\begin{center}
\psfrag{256}[c][c][.75][0]{$256$}
\psfrag{512}[c][c][.75][0]{$512$}
\psfrag{1024}[c][c][.75][0]{$1024$}
\psfrag{2048}[c][c][.75][0]{$2048$}
\psfrag{r}{$r$}
\includegraphics[width=8cm]{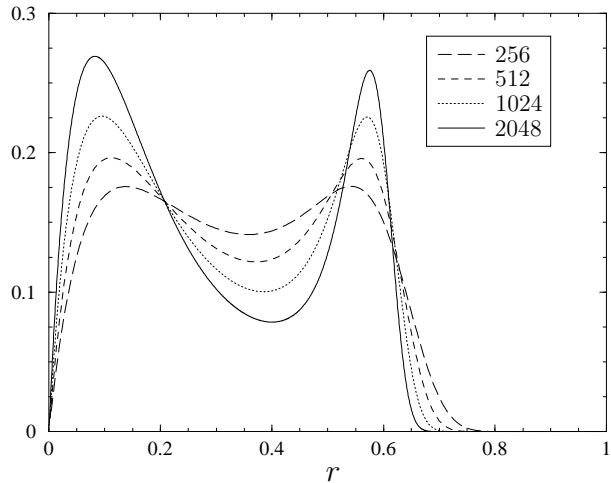}
\end{center}
\caption{\label{PcrN}
The probability distributions of the energy $r=M/N$ 
for pseudocritical values of $\kappa_{cr}(N)$ at $\beta=2.6$
for $N=256,512,1024,2048$.} 
\end{figure}

\section*{Finite size analysis results}

\subsection*{First order phase transition}

\begin{figure}
\begin{center}
\psfrag{22}[c][c][.75][0]{$2.2$}
\psfrag{24}[c][c][.75][0]{$2.4$}
\psfrag{26}[c][c][.75][0]{$2.6$}
\psfrag{28}[c][c][.75][0]{$2.8$}
\psfrag{30}[c][c][.75][0]{$3.0$}
\psfrag{r}{$r$}
\includegraphics[width=8cm]{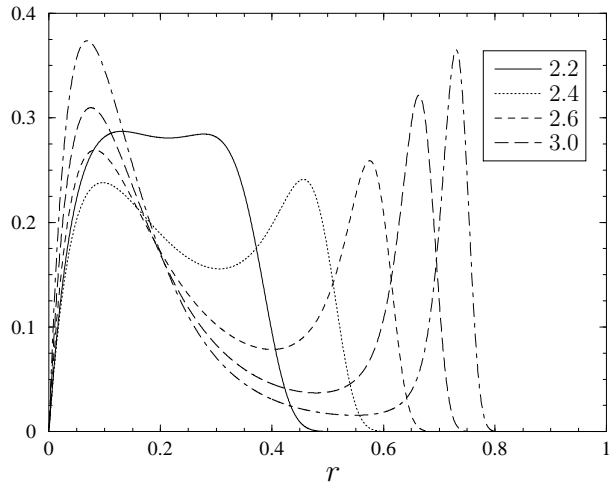}
\end{center}
\caption{\label{PcrB}
The probability distributions of the energy $r=M/N$ 
for pseudocritical values of $\kappa_{cr}(N)$ for $N=2048$
at $\beta=2.2, 2,4, 2,6, 2.8, 3.0$.} 
\end{figure}

We start the finite size analysis with the range $\beta>2$.
According to the discussion above,
the phase transition should be first order. 
A typical signal of first order phase transition
is a double peak in the distribution of the energy 
corresponding to the coexistence of two phases at the
transition with the latent heat being the separation
between peaks. Let us look for this signal in our case.
In figure~\ref{PcrN} we plot the pseudocritical
distributions of the energy density $r=M/N$ for 
different system sizes $N$ for $\beta=2.6$ for values
of $\kappa_{cr}(N)$ for which the both peaks have equal heights.
These values can be taken as pseudocritical ones.
The position of the left peak goes to zero when $N$ goes to
infinity, while of the right one to 
$r_{disc}(\beta)=\zeta(\beta)/\zeta(\beta-1) = 0.571 $ known
from the analytic calculations in the thermodynamic limit.
When $\beta$ changes the position of the right peak moves 
(figure~\ref{PcrB}). For large $\beta$, the
discontinuity $r_{disc}$ goes to one, while for $\beta \rightarrow 2$ 
it disappears and the two peaks merge. The depth of the
valley between the peaks increases with $N$ which means that the
configurations which do not belong to either of phases become more and
more suppressed.
The suppression becomes more visible for larger $\beta$ (figure~\ref{PcrB}).
On the contrary, for $\beta$ close to $2$ the valley is small or 
even absent. In this case the size of the system must be increased
sufficiently for the valley to  be visible.

\begin{figure}
\begin{center}
\psfrag{x}[c][c][.75][0]{$N^{-1}$}
\psfrag{y}[l][c][.75][-90]{$\kappa_{cr}(N)$}
\includegraphics[width=9cm]{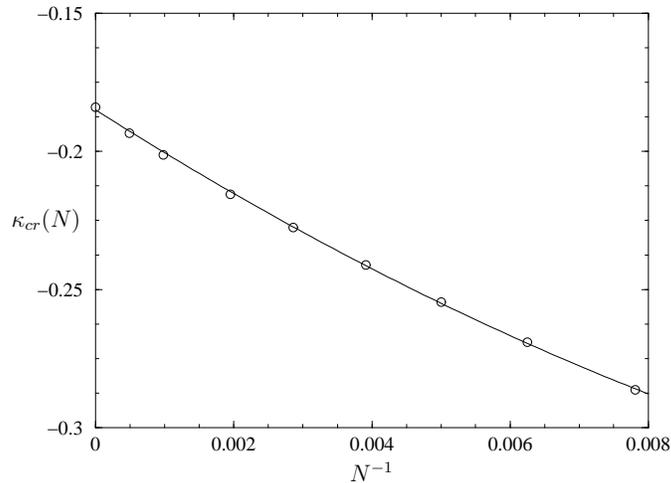}
\end{center}
\caption{\label{kN}
The pseudocritical value $\kappa_{cr}(N)$ versus $1/N$
for $\beta=3.0$. The curve going through the
data points corresponds to the best fit 
$\kappa=\kappa_{cr} + a/N \ (1+b/N)$.}
\end{figure}

The range of $\kappa$ for which 
two peaks coexist is called pseudocritical region.
For $\kappa$ in this region the relative heights of the peaks
vary. The extent of the region is inversely proportional to 
the size of the system, $N$. Outside the pseudocritical
region only one peak survives. If $\kappa$ changes within
the pseudocritical region, the average of the distribution
$r=\langle M \rangle/N$ moves very quickly between two peaks. The slope
of the curve $r(\kappa)$ grows linearly with $N$,
and eventually becomes infinite when $N$ is sent to infinity,
leading to the discontinuity $r_{disc}$ at $\kappa_{cr}$.
The slope of the curve corresponds to the second derivative,
which is heat capacity. Because the heat capacity grows linearly
with $N$, the system has latent heat corresponding to
the energy needed to move states of the system from one peak to the
other. Another characteristic signal of first order
phase transition which can be read off from the formula (\ref{fssn})
is that the position of the maximum of the second cumulant
should asymptotically lie on a curve $\Delta \kappa N = const$, which
means that $\kappa(N) = \kappa_{cr} + const/N$. 
As an example we show in figure~\ref{kN} the behaviour of
$\kappa_{cr}(N)$ versus $N$. The data points are fitted to the formula
with next-to-leading corrections of the standard form~: $\kappa(N) =
\kappa_{cr} + \frac{a}{N} (1 + \frac{b}{N})$.  The fit
gives\footnote{To estimate the errors of the fit parameters we use the
following procedure. The data consists of $n$ points for different
volumes.  We successively omit one of them and fit the formula to the
remaining ones obtaining $n$ different fits. Having done this, for
each parameter of the fit we have a distribution of $n$ values. We
find the average and the width of the distributions which we take as
the mean and the error of the parameters.  In doing this we assume
that all the data points are equally important, which is the simplest
possible assumption.  This procedure is not a statistical analysis,
but merely a way of presenting data.}  
$\kappa_{cr} = -0.185(1)$,
$a=-15.9(3)$ and $b=-24(2)$. The coefficient $b$ of the correction term
is large. Skipping the correction $b/N$ in the fit would decrease the
quality of the fit and  would significantly change the
estimate of the $\kappa_{cr}$ which agrees with the analytic
result $\kappa_{cr} = -\log \zeta(3) = -0.184$ with the correction term.

To summarize, the standard signals and finite size scaling 
characteristic of a first order transition 
are observed in the range $\beta\in (2,\infty)$ as expected.

\subsection*{Continuous phase transition}

\begin{figure}
\begin{center}
\psfrag{x}[c][c][.75][0]{$N$}
\psfrag{y}[l][c][.75][-90]{$$}
\includegraphics[width=9cm]{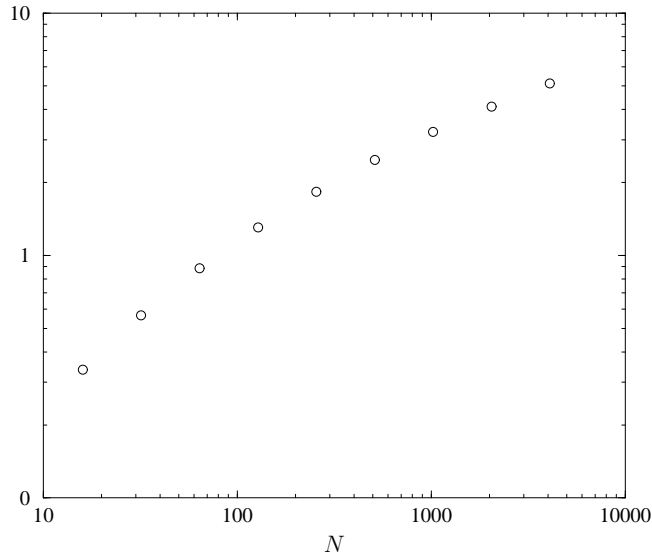}
\end{center}
\caption{\label{C3max}
The data points represent the maximum of the 
third cumulant for different $N$. They are plotted
in a log-log scale.} 
\end{figure}

\begin{figure}
\begin{center}
\psfrag{x}[c][c][.75][0]{$N$}
\psfrag{y}[l][c][.75][-90]{$A_3$}
\includegraphics[width=9cm]{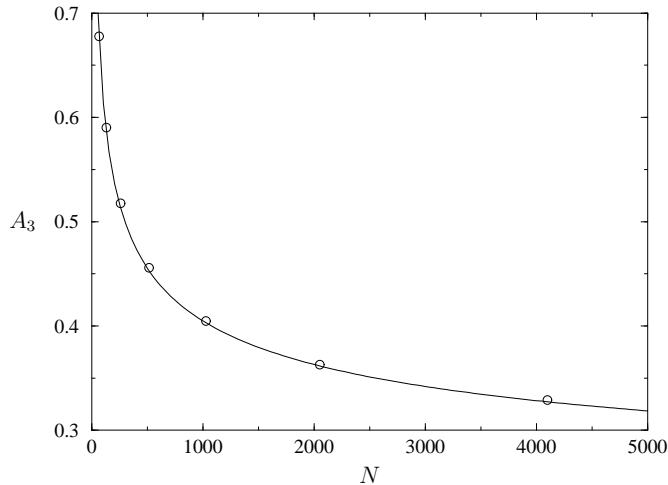}
\end{center}
\caption{\label{A3eff}
The effective exponent $A_3(N)$ calculated as a slope
of the lines in the previous figure.} 
\end{figure}

For $\beta \in (3/2,2)$ the linear growth of the maximum of the
second cumulant of the first order phase transition
changes to a sublinear behaviour $\sim N^{A_2}$,
where $0<A_2<1$, corresponding to a second order phase transition.
There is no double peak signal in the distribution of $M$.
The pseudocritical point is defined by the value of $\kappa$ at
which the second cumulant of the distribution of $M$ is
the largest. The second order phase transitions
have been extensively analyzed numerically in a number of papers. 
Therefore, we prefer here to go directly to the third order phase 
transition, $\beta \in (4/3,3/2)$, where some new ingredients 
like the presents of non-scaling extrema appear 
on top of the standard finite 
size effects known from second order transitions. 
As an example, we consider the case $\beta=1.4$ mentioned
before, for which we expect that
the second cumulant does not diverge, while the third
one does with the exponent $x_3 =-1/2$. As discussed before, we will
concentrate our attention on the leftmost maximum of the third
cumulant, which signals the appearance of the divergence in the
thermodynamic limit.  For $N$ going to infinity, the height of the
maximum is expected to grow as a power $N^{A_3}$ (\ref{fssn}),
possibly with some deviations for finite $N$ coming from non-scaling
part denoted by dots in (\ref{fssn}).  Indeed, it turns out that in
the range of $N$ from $16$ to $4096$ which we have covered by the
recursive method (\ref{recur}) the corrections to the asymptotic
formula are strong, as can be seen in figure~\ref{C3max}, where
the height of the maximum of the third cumulant versus $N$ in the
log-log scale is plotted. This is clearly non--linear and we define
the effective exponent $A_3(N)$ as the slope of the line fitted through
three consecutive points with $N$ standing for the biggest
size $N$ of the three points used in the linear fit. The effective
exponent is clearly far from reaching an asymptotic $N$--independent
value (figure~\ref{A3eff}).  To be able to find the asymptotic value
$A_3^* = A_3(N\rightarrow\infty)$ from this data, one would have to
know the form of subleading terms. Since we do not, we propose here a
phenomenological approach. We postulate a phenomenological form of
corrections to the running effective exponent 
\begin{equation}\label{A3fit}	
A_3(N) = A_3^* + b
N^{-c}.\end{equation}
It turns out, that this formula gives a good fit
(figure~\ref{A3eff}) with the values $A_3^* = 0.16(1)$, $b=1.58(1)$ and
$c=-0.27(1)$, estimated by the same procedure as discussed in the
footnote on one of earlier pages. From the fit one can make a
qualitative estimate of the subleading corrections. A simple
calculation shows that $N$ must be of order of $10^6$ in order that
the running exponent differ from the asymptotic value $A_3^*$, say, by
$0.04$. It is of course a rough estimate, but it points to how slowly
the corrections change with $N$.

\begin{figure}
\begin{center}
\psfrag{x}[c][c][.75][0]{$n$}
\psfrag{y}[l][c][.75][-90]{$A_n$}
\includegraphics[width=9cm]{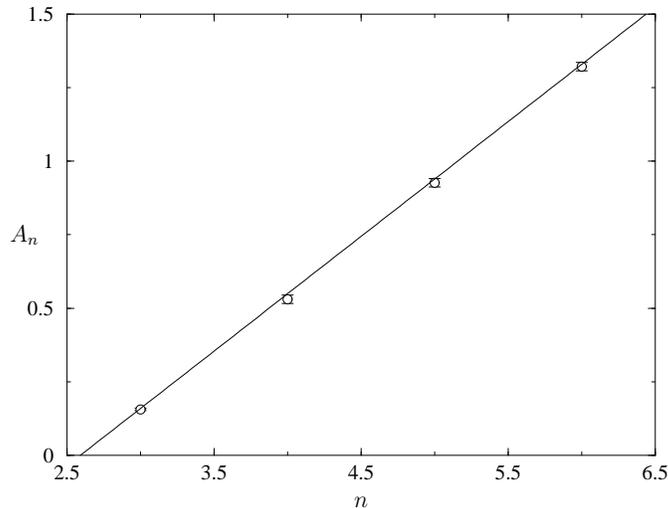}
\end{center}
\caption{\label{Bfit}
The linear fit of the exponents $A_n$ for $n=3,4,5,6$. 
The slope of the line corresponds to the exponent $B$.}
\end{figure}

The standard method of measuring the exponent $B$ relies on
tracing the position of the maximum for different sizes. 
For large $N$ it approaches an asymptotic value. This value
corresponds to the critical temperature. The exponent
in the function which measures the distance of the position
of the maximum from the critical temperature is proportional
to $\Delta \kappa \sim N^{-B}$. This gives the possibility of 
computng $B$. Due to the strong subleading corrections, 
this method does not give a good estimate in our case. 
Instead we propose another one. Unlike in the standard Monte 
Carlo, in our case the cost of computing cumulants does not 
grow with the order of the cumulant. We use this fact, to compute
the growth exponents $A_n$ for higher order -- divergent cumulants,
$n=3,4\dots$. Then we use the formula 
for linear growth~: $A_n = A_0 + n B$, which allows us to
compute the exponent $B$ as the slope of the line. 
As an input for $A_n$ we take
the asymptotic values $A_n^*$ with the errors, 
for $n=3,4,5,6$ estimated by the
method discussed above. The fit is shown in figure~\ref{Bfit}.
It gives~: $A_0=-1.005(42)$, $B=0.387(12)$ and $\chi^2/dof = 0.45$.
The errors quoted correspond to the 99.7\% confidence level\footnote{
For a cross-check we have also done a fit assuming equal errors 
on each $A_n$ and using condition $\xi^2/dof=1$.
The results
$A_0=-1.017(48)$ and $B=0.389(10)$ are consistent with the ones
obtained before.}.  The deviation from the 
theoretical values $A_0=-1.0$ and $B=0.4$ 
predicted from the Fisher relation are rather small~:
less than 1\% for $A_0$ and 4\% for $B$ . The former value
agrees within the errors with the theoretical one, while
the latter one is a little outside the error bars.

However, when one inserts those values to the formula $x_3 = -A_0/B-3$ 
and assumes the maximal correlation between $A_0$ and $B$, 
the errors are strongly enhanced and one obtains $x_3=-0.403(40)$. 
This is 20\% off from the theoretical value $x_3=-0.5$ 
and twice as much beyond the estimated errors. The assumption about the strong
correlation is dictated by the covariance matrix for the fit.
On the other hand, if we instead assume that the Fisher relation
indeed holds, as suggested by both the analytical and experimental 
evidence, we can set $A_0=-1$ and in this case we have only one
parameter relation $x_3 = 1/B - 3$ which leads to $x_3=-0.416(80)$. 

In interpreting the results one has to remember that they
depend on the phenomenological fit (\ref{A3fit}) extrapolated 
over several orders of magnitude. 
Such a fit gives meaningful results only due to the fact, 
that the small volume data points are {\em exact}, but on the
other hand one has to keep in mind that in general such 
a procedure may introduce a systematic error which is hard to quantify. 

Reduction of the error would require either a knowledge of the form
of the next-to-leading correction or pushing the computations to 
the system sizes of a few orders of magnitude larger.

\section*{Discussion}

The balls in boxes model provides a useful laboratory for testing
finite size scaling.  On one hand one can predict critical exponents
theoretically, on the other hand one can exactly calculate the finite
size partition function. In this respect the model is exceptional,
since most of the models must rely on noisy Monte Carlo data.
A comparison of the asymptotic form (\ref{fssn}) and the finite size
results indicates the presence of strong corrections to scaling. The
corrections come from the analytic and less singular part of the
partition function.  The corrections seem to be a very slowly varying
function of the system size as figure~\ref{A3eff} demonstrates. The
phenomenological approach employed to estimate the next--to--leading
corrections and the limiting asymptotic value of the effective
exponent is possible because unlike the Monte Carlo data, our data
points have no statistical uncertainties.

The finite size analysis results support the hypothesis that the
critical properties of the model are encoded in the one box
probability $\pi(q)$.  The universal scaling properties are obtained
from comparing two scales present in the model~: the inverse of an
exponential fall--off parameter of the effective probability
distribution $\pi(q)$ and the system size $N$.  The former one
measures the extent of fluctuations of the number balls in a box and
the latter one serves as a natural cut-off of those fluctuations.  As
mentioned in the introduction this type of transition plays an
important role in many physical systems.  In particular it is relevant
for understanding the transition in lattice gravity 
models\footnote{In lattice gravity the transition seems
to be generically of first order, and so far no continuous transition
has been observed.}.

The model provides a nice example of the finite size
pattern for a first order phase transition.
The results for the distribution are exact and easily
reproducible so one can measure all relevant quantities
and quickly  test any new ideas. Thus, we hope, the model may
also prove useful for studying generic properties of first order
transitions.

\section*{Acknowledgments}
We thank J. Jurkiewicz, A. Krzywicki and B. Petersson for discussions.
This work was partially supported by the Polish--British Joint
Research Collaboration Programme under the project WAR/992/137 and by
the French--German Joint Research Collaboration Programme PROCOPE
under the project 99/089.

P.B. was supported by the Alexander von Humboldt Foundation.  
L.B. thanks the August Witkowski Foundation for the scholarship.
D.A.J was partially supported by a Royal Society of Edinburgh/SOEID Support
Research Fellowship.

\end{document}